\documentclass[twocolumn,showpacs,preprintnumbers,amsmath,amssymb]{revtex4-1}

\usepackage{afterpage}
\usepackage[dvipdfmx]{graphicx}
\usepackage{color,bm}

\begin{document}

\title{
Earth rotation and 
time-domain reconstruction of polarization states for 
continuous gravitational waves from a known pulsar
} 
\author{Naoto Kuwahara}
\email{kuwahara@tap.st.hirosaki-u.ac.jp}
\author{Hideki Asada} 
\email{asada@hirosaki-u.ac.jp}
\affiliation{
${}^{ }$
Graduate School of Science and Technology, Hirosaki University,
Aomori 036-8561, Japan
}

\date{\today}

\begin{abstract} 
We consider effects of the Earth rotation on 
antenna patterns 
of a ground-based gravitational wave (GW) detector 
in a general metric theory 
that  allows at most six polarization states 
(two spin-0, two spin-1 and two spin-2) 
in a four-dimensional spacetime. 
By defining the cyclically averaged antenna matrix 
for continuous GWs from a known pulsar, 
we show that 
waveforms for each polarization state 
can be uniquely 
reconstructed in time domain 
from a given set of the strain outputs at a single detector. 
Constraining the propagation speed of 
extra polarization modes, if they 
coexist with the transverse-traceless modes,  
is also discussed. 
We examine also possible effects 
due to the length-of-day modulation  
as well as a secular change in the pulsar spin period. 
\end{abstract}


\maketitle

\section{Introduction}
A century after the birth of Einstein's theory of general relativity (GR) 
\cite{Einstein1916, Einstein1918}, 
the first direct observation of gravitational waves (GWs) was done 
for the golden event GW150914. 

GR is not perfectly consistent with quantum physics and 
string theoretical viewpoints.  
It is thus important to probe new physics beyond GR 
\cite{Will, LVK, KAGRA}. 
In a four-dimensional spacetime, 
general metric theories  
allow at most six GW polarization states 
(two spin-0, two spin-1 and two spin-2) 
\cite{Eardley}. 
Once the 
transverse-traceless (TT) 
polarizations are detected, 
it will be of great importance to probe the extra polarizations beyond GR.  
The two scalar modes called Breathing (B) and Longitude (L)  
are degenerate in interferometry, 
because the antenna pattern functions for  B and L modes 
take the same form but with the opposite sign 
\cite{Nishizawa2009}.  
Therefore, a direct test of each polarization state needs 
five or more ground-based detectors. 
For a merger event associated with an electromagnetic counterpart, 
we can know the GW source sky position by the multi-messenger astronomy. 
For such multi-messenger events in particular sky regions, 
the minimum requirement 
becomes four ground-based detectors including KAGRA 
\cite{Hagihara2018, Hagihara2019, Hagihara2020, PTEP, KAGRA-2022}.

The GW150914 data fits well with a binary black hole merger in GR 
\cite{LIGO2016}, 
though this test is inconclusive 
because the number of GW polarization states in GR 
is equal to the number of aLIGO detectors. 
The addition of Virgo to the aLIGO detectors for GW170814 
enabled the first informative test of GW polarizations. 
According to their analysis, 
the GW data are described much better by the pure tensor modes 
than pure scalar or pure vector modes 
 \cite{LIGO2017}. 
A range of tests of GR for GW170817, 
the first observation of GWs from 
a binary neutron star inspiral  \cite{GW170817}, 
were done by aLIGO and Virgo \cite{LIGO2019}. 
The tests include a test similar to Ref. \cite{LIGO2017} by performing 
a Bayesian analysis of the signal properties with the three detector outputs, 
using the tensor, the vector or the scalar response functions, 
though the signal-to-noise ratio in Virgo was much lower than those 
in the two aLIGO detectors. 
The prospects for polarization tests were discussed (e.g. \cite{Hayama, Isi2015,Isi2017,Takeda}).

GW signals are a linear combination of 
different polarization modes, 
where the coefficients of each mode is called the antenna pattern function 
that depends on the polarization state as well as the source direction 
\cite{ST, GT, CYC, PW, Book-Creighton, Book-Maggiore}. 
For a merger event so far, 
the antenna pattern is almost instantaneous. 
As a result, the required minimum number of detectors 
must equal to the number of independent polarization states 
when we wish a direct separation of all the possible polarizations states.

It is thus interesting to search continuous GWs from pulsars 
\cite{Jaranowski}. 
There are three types of continuous GWs searches. 
Targeted searches look for signals from known pulsars,  
for which the spin periods can be accurately determined 
mainly from radio observations 
\cite{LIGO-ApJ-2017, LIGO-pulsar-2017, LIGO-PRL-2018, LIGO-pulsar-2019, LIGO-PRD-2019, LIGO-2111, LIGO-2112, LIGO-PRD-2022}. 
Directed searches look for signals from known sky locations
\cite{LIGO-ApJ-2021b, LIGO-PRD-2022b, LIGO-2201b, LIGO-2204b}. 
All-sky searches look for GW signals from unknown sources  
\cite{Papa2019, Papa2021, LIGO-PRD-2021, LIGO-best-2022, Papa2202b}. 
From LIGO and Virgo O3 data, 
for instance, 
the best upper limits 
on the GW strain amplitude 
for all-sky search 
have been recently obtained as 
$\sim 1 \times 10^{-25}$ in the frequency range of 100 to 200 Hz 
\cite{LIGO-best-2022}. 
In addition, there are all-sky searches also 
for unknown neutron stars in binary systems 
\cite{PRD-2021c, ApJ-2022c}.

Besides a direct search of 
mixed non-GR polarization states,  
there exists the binary pulsar test 
which is a comprehensive study 
of the orbital decay. 
From the orbital decay observation of 
the binary pulsar B1913+16, 
the radiation flux by extra polarizations 
has been limited to less than $\sim 0.1\%$ 
\cite{Will, Weisberg2016}.  
Very recently, 
Kramer et al. have reported that 
the double pulsar PSR J0737–3039A/B 
validates the prediction of GR 
more precisely at the level of $\sim 1 \times 10^{-4}$ 
\cite{Kramer2021}. 
These binary/double pulsar observations imply 
that the gravitational radiation due to non-GR polarizations 
must be much weaker than that of GR ones, 
even if they coexist. 
It is thus important to discuss a GW data analysis method 
for searching such a small amplitude of 
non-GR polarizations, if they coexist with GR ones.

In pioneering work \cite{Isi2015, Isi2017}, 
Isi and his collaborators developed a method 
that allows to separate the non-GR 
as well as GR polarizations for continuous GWs 
by taking account of the Earth rotation. 
In their work, 
each polarization is sinusoidal 
with fitting parameters.


Does the Earth rotation allow to reconstruct 
a time-domain waveform 
of GW polarization states for a known pulsar?   
It is an open issue whether 
non-GR waveforms in time domain are sinusoidal, 
because we do not currently know the true theory of gravity. 
In expectation of a sensitivity significantly improved 
by the future third-generation detectors such as 
the Cosmic Explorer (CE) and the Einstein Telescope (ET) 
\cite{ET, CE, CE2}, 
the main purpose of the present paper is 
to demonstrate that 
the Earth rotation 
allows to reconstruct waveforms in time domain 
for each polarization state of the pulsar GWs, 
if non-GR polarizations exist, 
where any GW template is not assumed a priori 
except for being periodic.

This paper is organized as follows. 
Section II briefly summarizes 
expressions for 
the antenna pattern functions 
and the strain outputs. 
Section III discusses the cyclically averaging of the antenna patterns 
in order to demonstrate the 
time-domain reconstruction of 
each polarization state. 
Section VI mentions future prospects along the direction of this study 
and possible other effects. 
Section V is devoted to Conclusion. 


\section{Antenna patterns and GW signals}
In a four-dimensional spacetime, 
a general metric theory allows 
six polarizations at most \cite{Eardley}; 
$h_B(t)$ for the spin-0 B mode, 
$h_L(t)$ for the spin-0 L mode, 
$h_V(t)$ and $h_W(t)$ for two spin-1 modes, 
$h_+(t)$ for the plus mode 
and 
$h_{\times}(t)$ for the cross mode. 
For a laser interferometer, 
the antenna pattern function to each polarization 
is denoted as 
$F^{I}(t)$, where $I = B, L, V, W, +, \times$ 
\cite{PW}. 
It depends on the GW source direction $\theta$ and $\phi$ 
as well as the polarization angle $\psi$. 
The latitude and longitude of a GW source 
are functions of time $\theta(t)$ and $\phi(t)$, 
whereas they are almost instantaneous  
for a merger or burst event.  
The change of the detector arm directions with time 
is also taken into account 
when calculating the antenna pattern functions 
through $\psi(t)$ 
\cite{Isi2015, Isi2017, Takeda, Takeda2019}. 
For the brevity, 
we use only $t$ in the notation of the antenna pattern.

The strain output at the detector 
is written as \cite{Nishizawa2009, Isi2015, Isi2017, Takeda, Takeda2019, PW, Book-Creighton, Book-Maggiore}
\begin{align}
S(t) =& F^S(t) h_S(t) + F^V(t) h_V(t) + F^W(t) h_W (t) 
\notag\\
&+ F^+(t) h_+(t) + F^{\times}(t) h_{\times}(t) 
+ n(t)
\nonumber\\
=& \sum_{I = S, V, W, +, \times} 
F^I(t) h_I(t) + n(t) ,  
\label{Sa}
\end{align}
where we define $F^S(t) \equiv F^B(t) = - F^L(t)$, 
we denote $h_S(t) \equiv h_B(t) - h_L(t)$, 
and $n(t)$ means noises. 
In the rest of this paper, $I \in S, V, W, +, \times$ 
is denoted simply as $I$. 

For LIGO-Virgo merger events, 
the duration is roughly 
$\sim 1-1000$ milliseconds ($\ll T_E$), 
where $T_E$ is the Earth rotation period $\sim 24$ hours.  
The time variation of $F^I(t)$ is negligible enough for us to safely use the instantaneous antenna pattern 
for the data analysis. 
The dependence on time 
is discussed e.g. \cite{Isi2015, Isi2017, Takeda2019}. 

On the other hand, 
the antenna pattern 
changes significant with time in a day.

\section{Time-domain Reconstruction for periodic GWs}
\subsection{$N$-cycle Averaging} 
We consider periodic GWs with period $T_P$ 
as 
\begin{align}
h_I(t) = h_I(t+n T_P) , 
\label{hI} 
\end{align}
where $n$ is an integer. 
It is sufficient to consider 
$h_I(t)$ only for $t \in [0, T_P)$ 
because of being periodic.

\begin{figure}
\includegraphics[width=7.5cm]{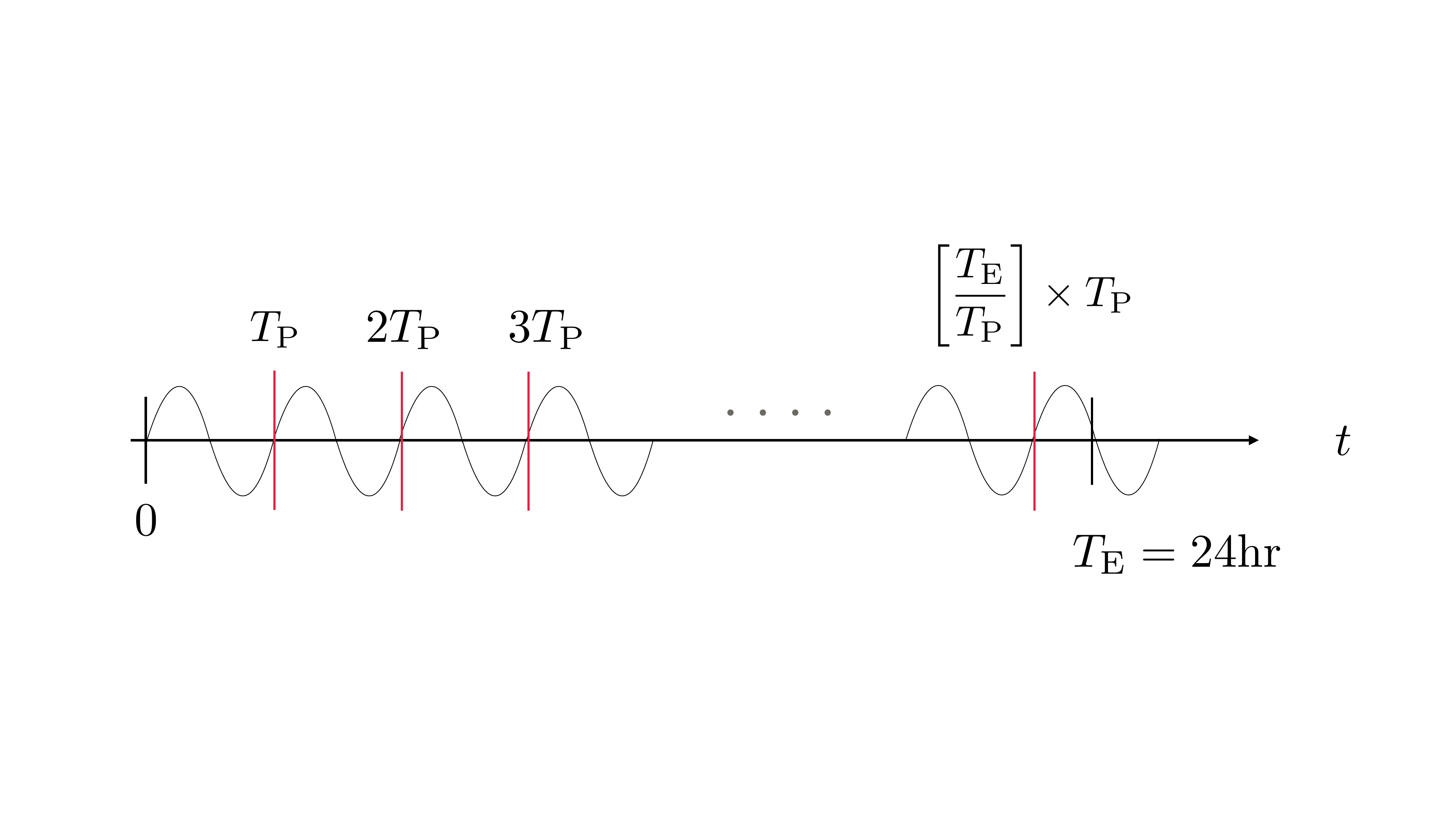}
\caption{
Schematic figure for each cycle of periodic GWs. 
}
\label{figure-period}
\end{figure}

\begin{figure}
\includegraphics[width=7.5cm]{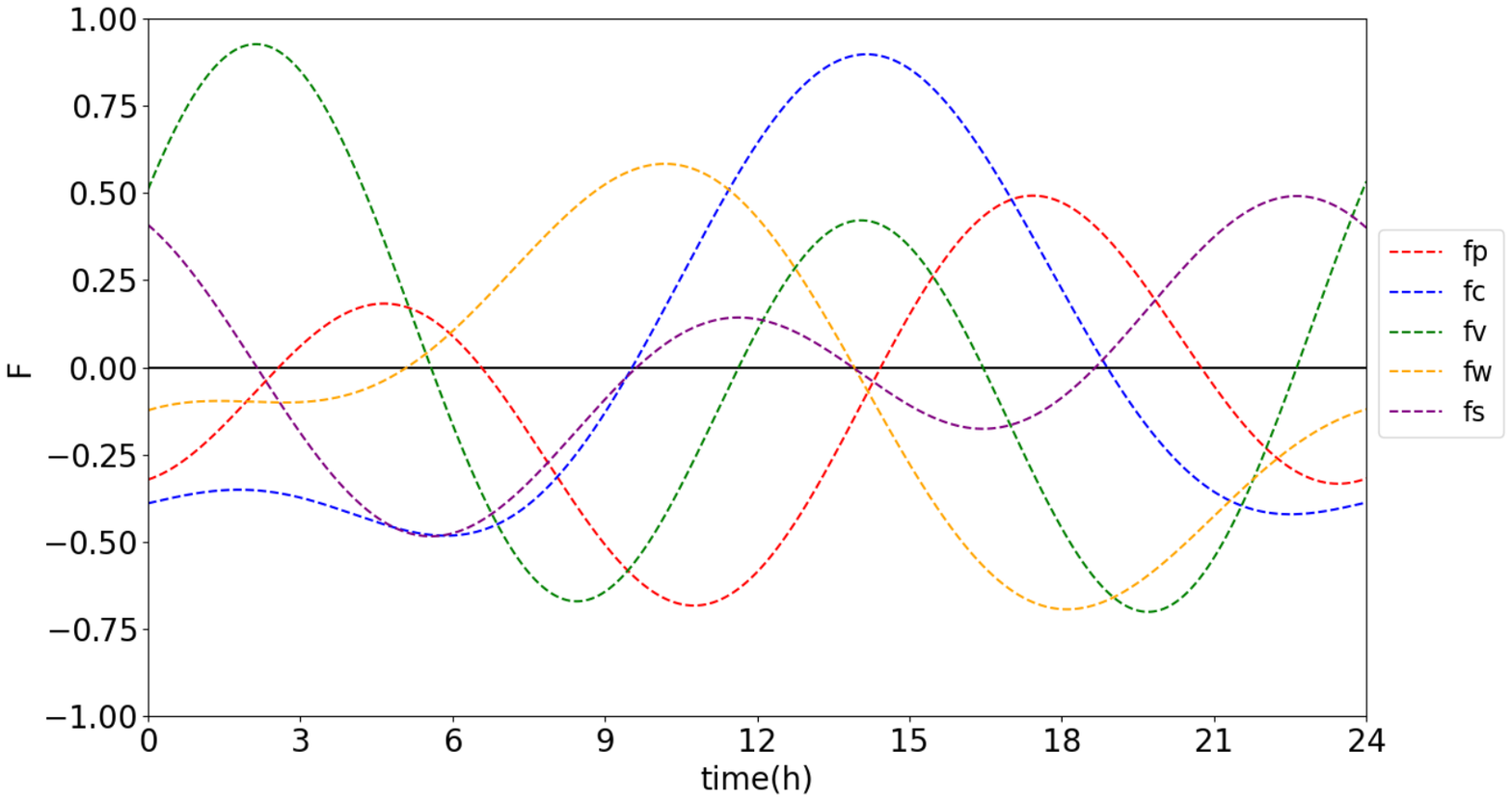}
\caption{
Daily variation in the antenna patterns for each GW polarization. 
For its simplicity, 
the location of the LIGO-Hanford detector and its arm direction 
are assumed for the sky location of the Crab pulsar. 
}
\label{figure-F}
\end{figure}

For the sake of simplicity, 
we focus on one day as the observational duration, 
where the number of the GW cycles in one day 
is 
$N_E \equiv [T_E/T_P]$ 
for the Gauss symbol $[ \: ]$, namely the integer part 
as shown by Figure \ref{figure-period}. 
Note that $h(t)$ is cyclic with period $T_P$, 
while $F_I(t)$ has another period $T_E$. 
See Figure \ref{figure-F} for a daily variation of 
$F_I(t)$ for each polarization.

For $N$ cycles, the 
strain outputs can be expressed in terms of 
the periodic function $h_I(t)$ 
and stochastic $n(t)$. 
We divide the total $N$ cycles into each one cycle 
of $t \in [(a-1) T_p, a T_p)$, 
where $a = 1, 2, \cdots, N$ is an integer. 

The strain output in 
the $a$-th cycle is  
denoted as 
$S_a(t) \equiv S(t + (a-1) T_P)$ 
for $t \in [0, T_p)$, 
which is written explicitly as 
\begin{align}
  S_{1}(t) \equiv& S(t) 
  \notag\\
  =& \sum_I F^I(t) h_I(t) + n(t) ,
  \notag \\
  S_{2}(t) \equiv& S(t+T_P) 
  \notag\\
  =& 
  \sum_I F^I(t+T_P) h_I(t+T_P) + n(t+T_P) ,
  \notag \\
  \cdots& 
\notag \\
  S_{N}(t) \equiv& S(t+(N-1)T_P) 
  \notag\\
  =& \sum_I F^I(t+(N-1)T_P) 
  h_I(t+(N-1)T_P)
  \notag\\
  &~~~~~ + n(t+(N-1)T_P) .
  \label{S}
\end{align}
Note that $S_a(t) \neq S_b(t)$ for $a \neq b$, 
because $F^I(t)$ changes periodically with the Earth rotation 
but its period is not $T_p$ but $T_E$. 


In order to use the least square method, therefore, 
let us define $A(t)$ by 
\begin{align}
  A(t) \equiv& 
  \left(S_{1}(t) - \sum_I F^I_{1}(t) h_I(t)\right)^{2} 
  \notag\\
  &+ \cdots + \left(S_{N}(t) - \sum_I F^I_{N}(t) h_I(t)\right)^{2} 
  \notag \\
  =& 
  \sum_{a=1}^{N} \Bigl(S_{a}(t) - F^I_a(t) h_I(t)\Bigr)^{2} ,
  \label{A}
\end{align}
where Eqs. (\ref{hI}) and (\ref{S}) are used 
and $F^I_a(t) \equiv F^I(t + (a-1) T_P)$.
 In the rest of the paper, 
the $N$-cycle sum $\sum\limits_{a=1}^N$ 
is abbreviated as $\sum\limits_a$.

In the least square method, 
the expected $h_I(t)$  at time $t$, 
denoted as $h_{IN}(t)$, 
is determined by five equations as 
$\partial A(t)/\partial h_I(t) = 0$ for each $I$, 
where the subscript $N$ indicates the dependence on 
the number of cycles. 
Note that $h_{IN}(t) \neq h_I(t)$, 
because $h_{IN}(t)$ depends on the number of the cycles. 
According to the laws of large numbers in probability theory, 
$h_{IN}(t)$ approaches the true $h_I(t)$ 
as $N \to \infty$.

The coupled equations for 
$h_{IN}(t)$  
are rearranged in a vectorial form as 
\begin{align}
M_N(t) \vec{H}_N(t) = \vec{L}_N(t) ,
\label{eq-vec}
\end{align}
where we define 
\begin{widetext}
\begin{align}
\overrightarrow{H}_N(t) &\equiv
  \begin{pmatrix}
    h_{+ N}(t) \\
    h_{\times N}(t) \\
    h_{V N}(t) \\
    h_{W N}(t) \\
    h_{S N}(t)
  \end{pmatrix} , 
  \label{H}
  \\
%
\overrightarrow{L}_N(t) &\equiv  
\begin{pmatrix}
    \sum\limits_a F^{+}_{a}(t) S_{a}(t) \\
    \sum\limits_a F^{\times}_{a}(t) S_{a}(t) \\
    \sum\limits_a F^{\rm{V}}_{a}(t) S_{a}(t) \\
    \sum\limits_a F^{\rm{W}}_{a}(t) S_{a}(t) \\
    \sum\limits_a F^{\rm{S}}_{a}(t) S_{a}(t)
  \end{pmatrix} ,
  \label{L}
\\
%
%
M_N(t) &\equiv 
\begin{pmatrix}
    \sum\limits_a [F^{+}_{a}(t)]^{2} & \sum\limits_a F^{+}_{a}(t) F^{\times}_{a}(t) 
    & \sum\limits_a F^{+}_{a}(t) F^{V}_{a}(t) & \sum\limits_a F^{+}_{a}(t) F^{W}_{a}(t) 
    & \sum\limits_a F^{+}_{a}(t) F^{S}_{a}(t) \\
    \sum\limits_a F^{\times}_{a}(t) F^{+}_{a}(t) & \sum\limits_a [F^{\times}_{a}(t)]^{2} 
    & \sum\limits_a F^{\times}_{a}(t) F^{V}_{a}(t) & \sum\limits_a F^{\times}_{a}(t) F^{W}_{a}(t) 
    & \sum\limits_a F^{\times}_{a}(t) F^{S}_{a}(t) \\
    \sum\limits_a F^{V}_{a}(t) F^{+}_{a}(t) & \sum\limits_a F^{V}_{a}(t) F^{\times}_{a}(t) 
    & \sum\limits_a [F^{V}_{a}(t)]^{2} & \sum\limits_a F^{V}_{a}(t) F^{W}_{a}(t) 
    & \sum\limits_a F^{V}_{a}(t) F^{S}_{a}(t) \\
    \sum\limits_a F^{W}_{a}(t) F^{+}_{a}(t) & \sum\limits_a F^{W}_{a}(t) F^{\times}_{a}(t) 
    & \sum\limits_a F^{W}_{a}(t) F^{V}_{a}(t) & \sum\limits_a [F^{W}_{a}(t)]^{2} 
    & \sum\limits_a F^{W}_{a}(t) F^{S}_{a}(t) \\
    \sum\limits_a F^{S}_{a}(t) F^{+}_{a}(t) & \sum\limits_a F^{S}_{a}(t) F^{\times}_{a}(t) 
    & \sum\limits_a F^{S}_{a}(t) F^{V}_{a}(t) & \sum\limits_a F^{S}_{a}(t) F^{W}_{a}(t) 
    & \sum\limits_a [F^{S}_{a}(t)]^{2}
  \end{pmatrix} .
  \label{M}
\end{align}
\end{widetext}

The solution for $h_{IN}(t)$ is thus 
\begin{align}
\overrightarrow{H}_N(t) = M_N^{-1}(t) \overrightarrow{L}_N(t) ,
  \label{sol}
\end{align}
where $M_N^{-1}(t)$ is the inverse matrix of $M_N(t)$. 
$\overrightarrow{L}_N(t)$ in the right-hand side of Eq. (\ref{sol}) 
includes noise through $S_a(t)$. 
Thereby the reconstructed waveform of $\overrightarrow{H}_N(t)$ 
($\neq \overrightarrow{H}(t)$) 
is influenced by noise.

We refer to $M_N(t)/N$ as 
the cyclically averaged antenna matrix (CAAM), 
because the procedure of $\frac1N \sum_a$ is 
the averaging for the $N$ cycles. 
One may ask if $M(t)/N$ corresponds to the covariance matrix. 
This is not the case, 
because the averaging of $F^I(t)$ as $\frac1N \sum_a F^I_a(t)$ 
does not always vanish.

The formal solution as Eq. (\ref{sol}) with Eqs. (\ref{H})-(\ref{M}) 
shows clearly the existence and uniqueness 
of the solution for the inverse problem. 
In practical calculations, however, 
we do not need obtain $M^{-1}(t)$, 
for which numerically performing 
the inverse of a matrix is rather time-consuming. 
It is sufficient and even convenient 
to solve Eq. (\ref{eq-vec}) by using 
a more sophisticated algorithm,  
e.g. LU (lower-upper) decomposition 
\cite{NumericalRecipe}, 
which makes numerical calculations faster than the inverse matrix method.


\begin{figure*}
\includegraphics[width=8.5cm]{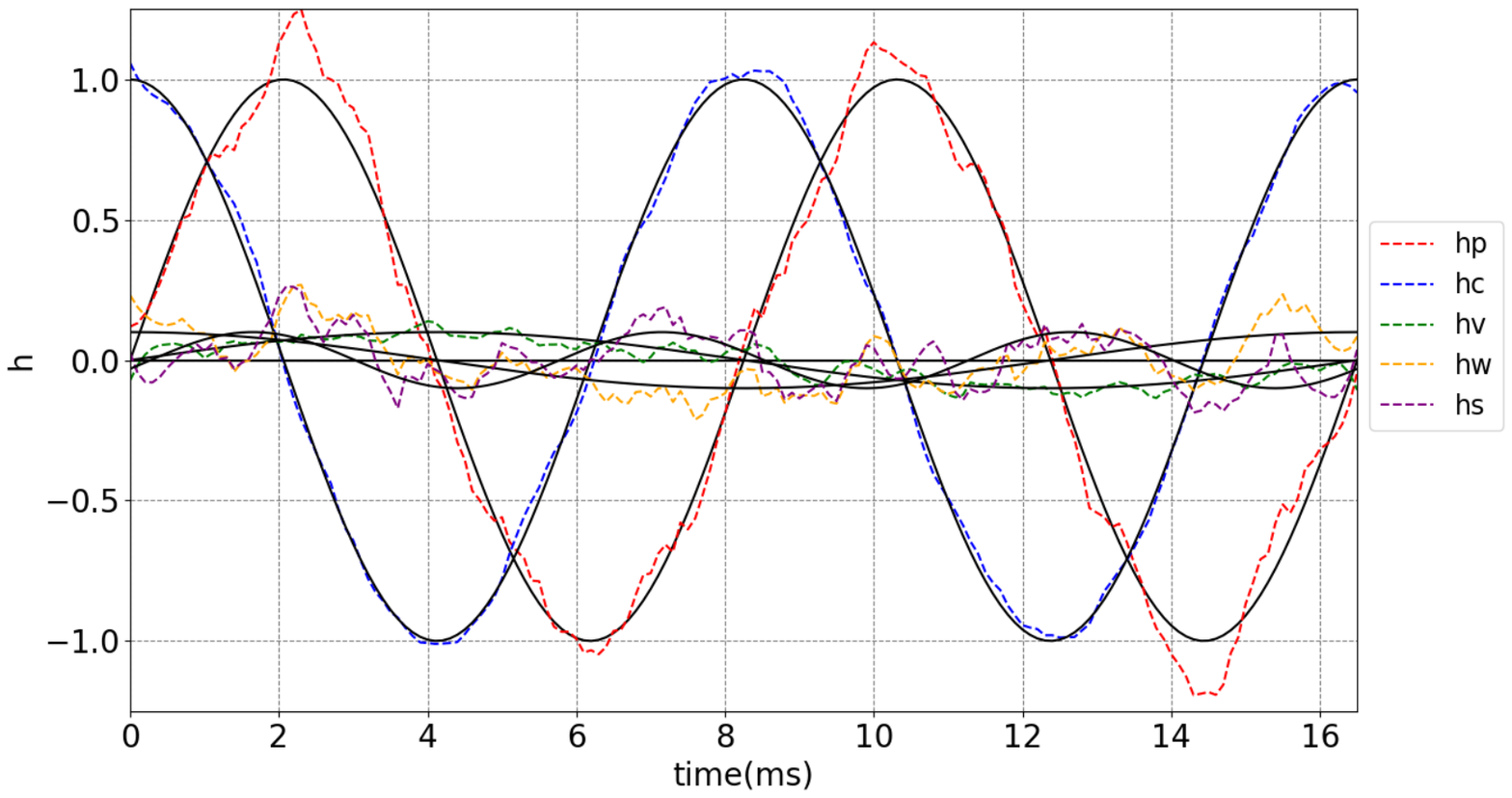}
\includegraphics[width=8.5cm]{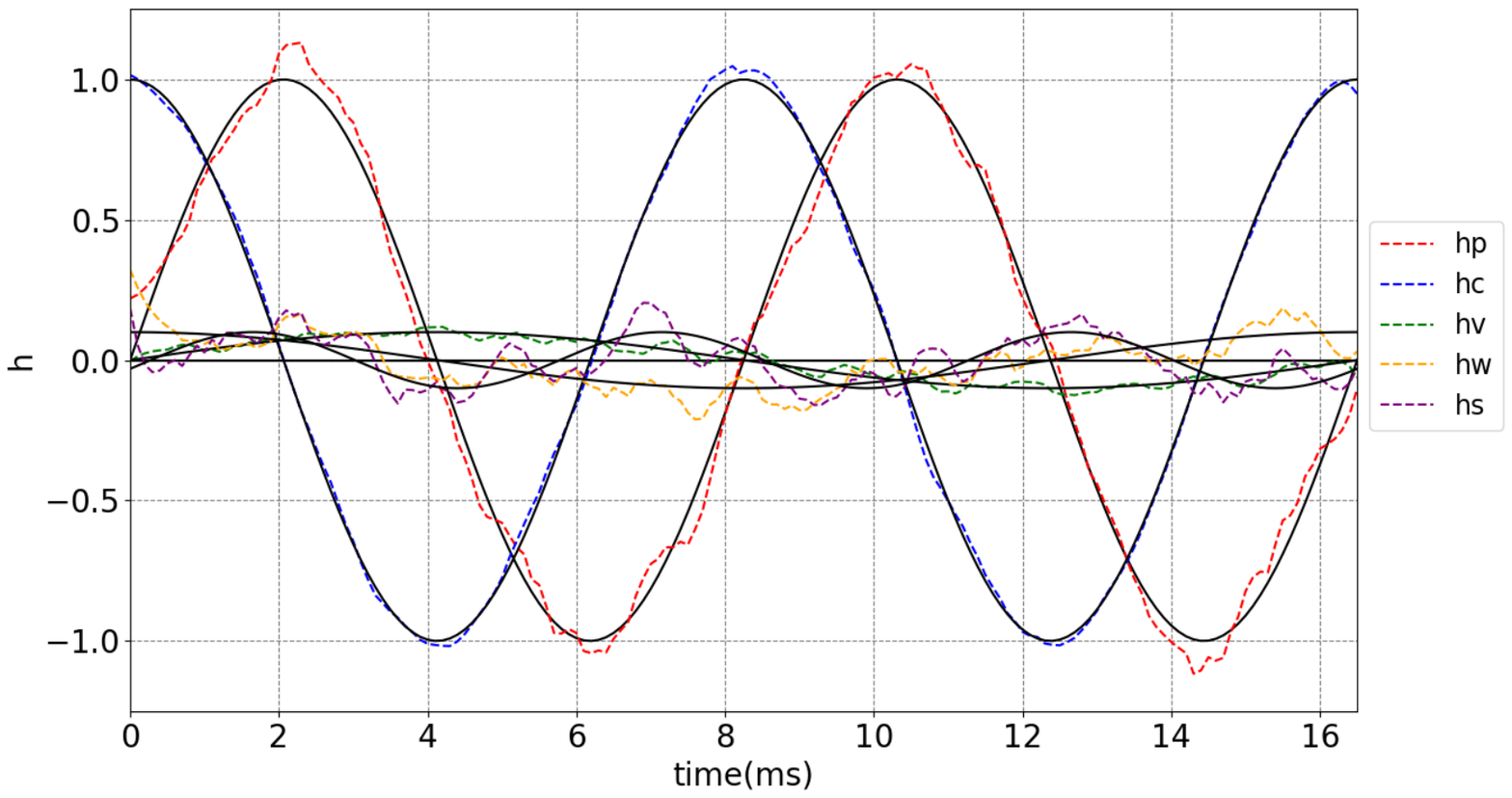}
\includegraphics[width=8.5cm]{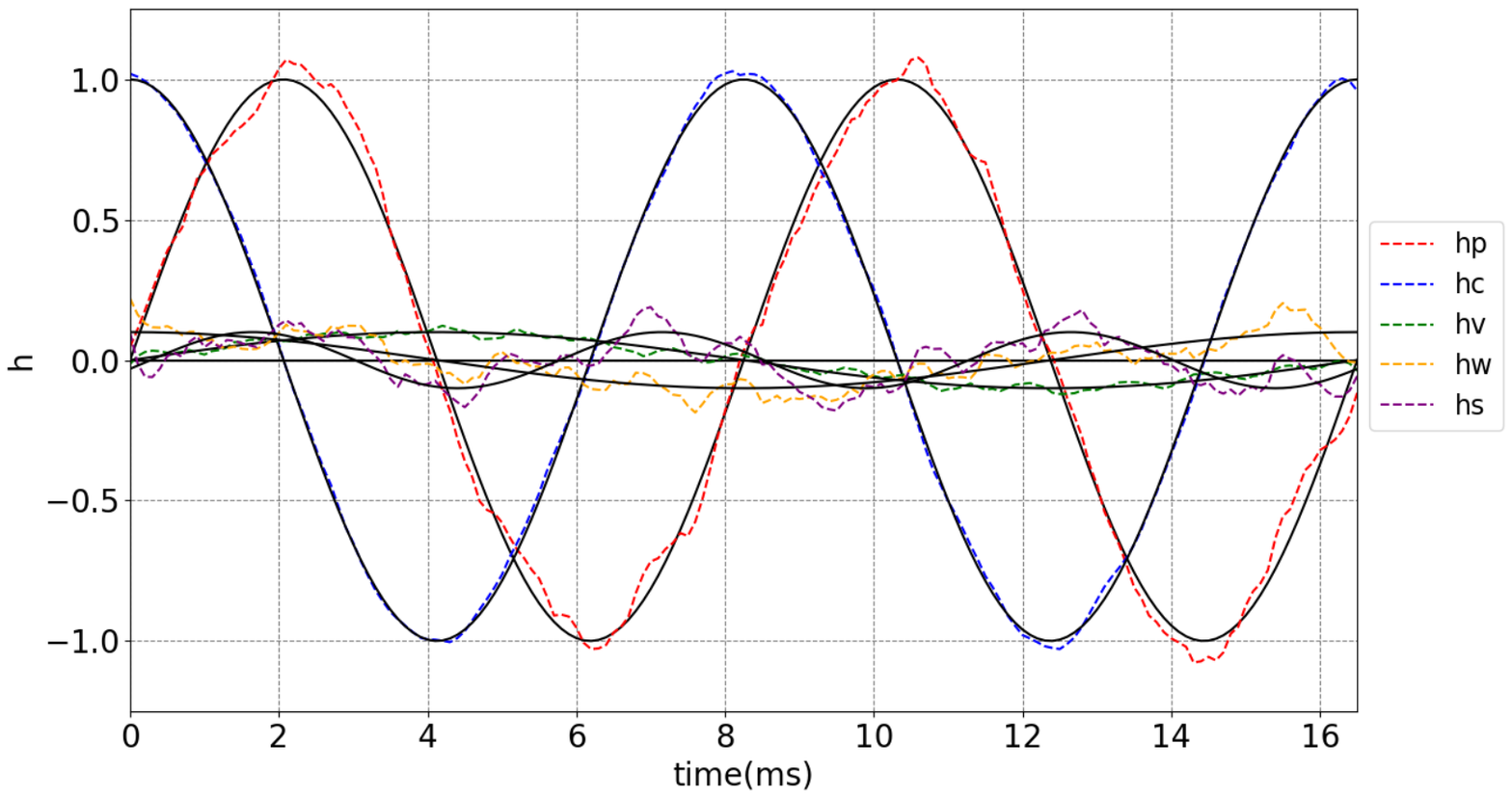}
\includegraphics[width=8.5cm]{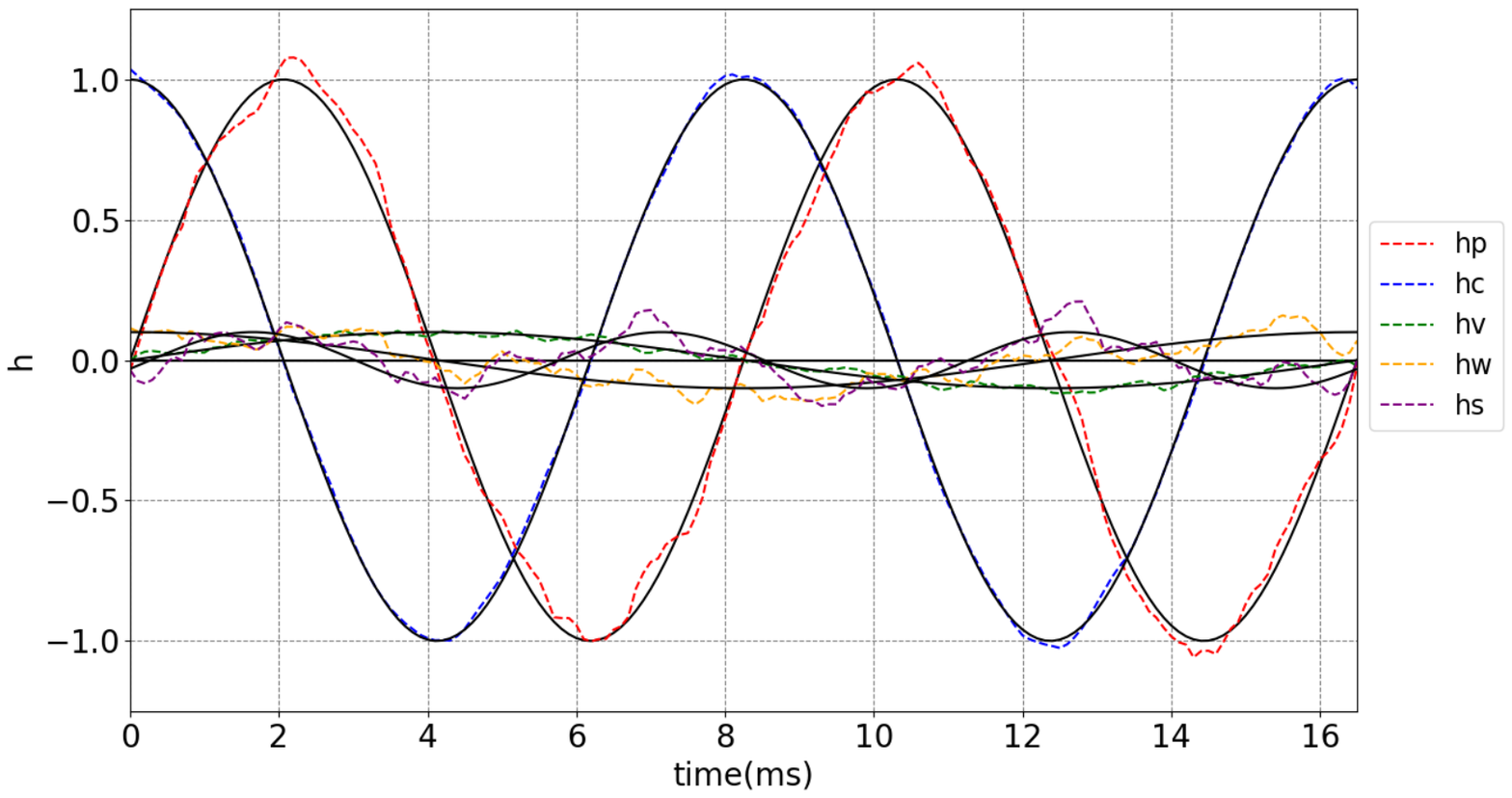}
\caption{
Time-domain reconstruction: 
From $S(t)$ to $h_IN(t)$ by Eq. (\ref{sol}). 
The uint of the vertical axis is arbitrary. 
Top left: 
$N = 1309090$, 
Top right: 
$N =  2618181$, 
Bottom left:  
$N =  3927272$, 
Bottom right: 
$N = 5236363$, 
each of which  corresponds to 
$6, 12, 18, 24$ hours, respectively. 
The LIGO-Hanford detector  
configuration (its position and arm direction) 
and the Crab pulsar 
($T_p = 16.5$ msec.) are assumed, 
where 
GW waveforms follow sine functions, 
indicated by solid black (in color) lines. 
For exaggerations, 
the GW amplitude for the extra polarizations (S, V, W) 
is chosen as 0.1, 
and 
the noise $n(t)$ obeys a Gaussian distribution 
with the standard deviation of 20, 
such that the plots can be recognized by eyes.  
For $N = 1309090$ cycles ($6$ hours), 
the TT modes 
are well reconstructed, 
whereas 
the S, V, W modes and noises are hardly distinguishable from each other 
by eyes. 
As $N$ increases, 
the noise is effectively reduced as 
$n_{eff}(t) \propto  1/\sqrt{N}$. 
The S, V and W modes 
are thus reconstructed in time domain better 
for $N =$5236363 (24 hours). 
As a simple example, 
an offset is considered only for $h_S(t)$, 
which may reflect the arrival time difference 
due to the polarization-dependent speed of gravity. 
The arrival time delay is chosen as $T_p/60$. 
The offset is reconstructed in the present method 
using Eq. (\ref{sol}).
}
\label{figure-h}
\end{figure*}

\begin{figure}
\includegraphics[width=8.5cm]{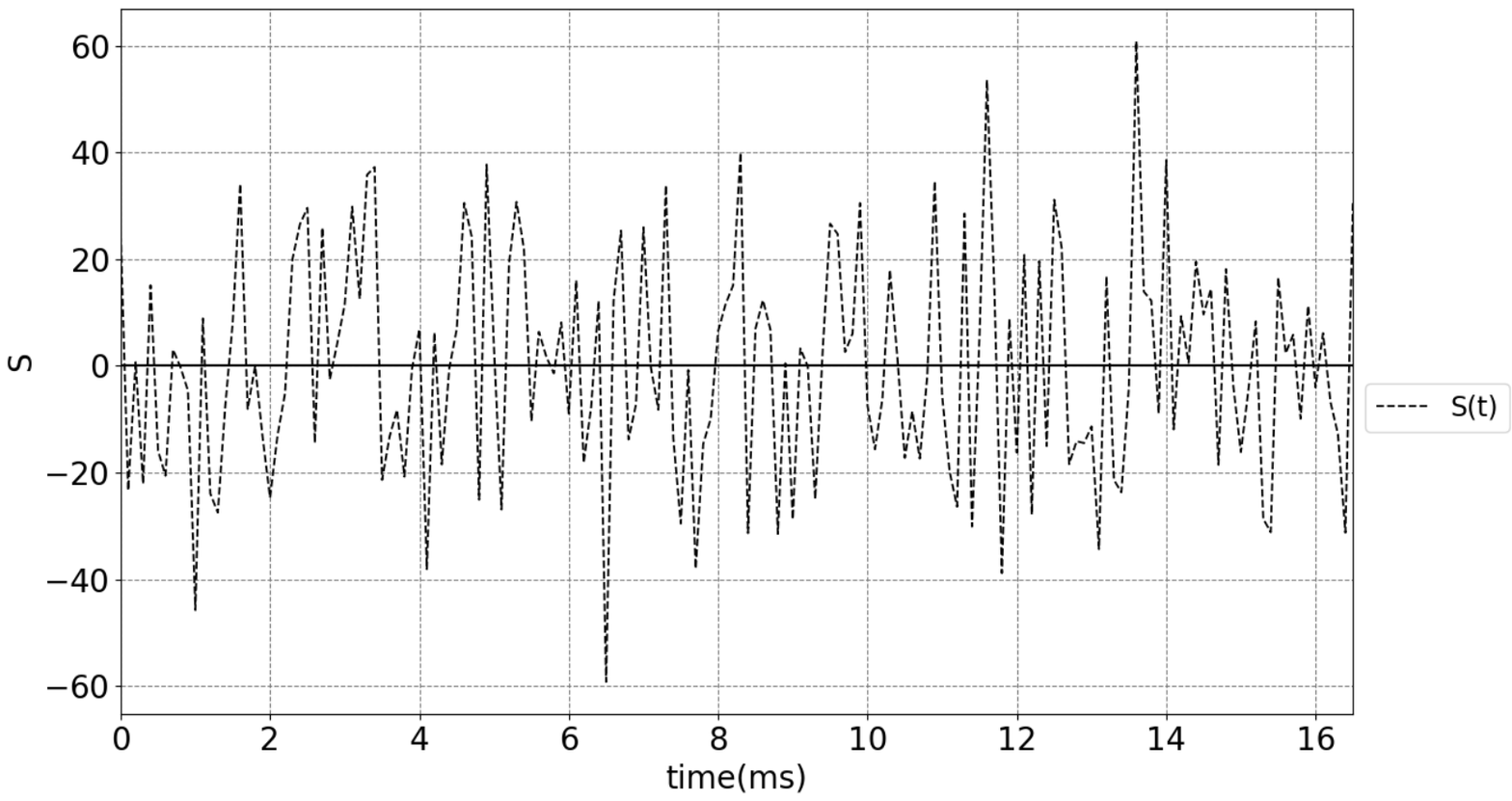}
\caption{
Mock data of strain outputs for one of the $N$ cycles 
(16.5 msec.) in the numerical calculations 
for Figure \ref{figure-h}. 
The standard deviation of the noise is chosen as 20, 
where  the amplitude of TT modes is the unity. 
One cannot recognize the TT signal only from this plot. 
}
\label{figure-data}
\end{figure}

Up to this point, 
we have assumed $\det M_N \neq 0$, 
where $\det$ denotes the determinant of a matrix. 
If $\det M_N = 0$, 
it describes a curve in the sky, 
for which the CAAM is degenerate.
Does $\det M_N = 0$ occur? 
No, $\det M_N$ does not vanish. 
It is positive anywhere in the sky, 
according to numerical computations. 
Currently, a mathematical proof of $\det M_N > 0$ 
is not obtained, because the expression of 
$\det M_N$ is very complicated.

\subsection{Numerical examples}
Figure \ref{figure-h} 
shows numerical 
time-domain reconstructions of  
waveforms for each polarization state,   
where one day observation 
and a pulsar GW period of 
16.5 milliseconds 
(e.g. for the Crab pulsar) 
are assumed,  
corresponding to 
$N \sim 5 \times 10^6$,  
because of the limited computational resources. 
In this figure, 
the amplitudes of $h_+(t)$ and $h_{\times}(t)$
are equal to each other,  denoted simply as $h_{TT}$. 
$h_S = h_V = h_W = h_{TT}/10$ 
and $\bar{n} = 20 \times h_{TT}$ 
are chosen for exaggerations, 
such that plots can be recognized by eyes. 
Here, the amplitudes of $S$, $V$ and $W$ modes 
are denoted as $h_S, h_V, h_W$, respectively, 
and the standard deviation of the noise is 
denoted as $\bar n$. 
See Figure \ref{figure-data} for 
strain outputs during one of the $N$ cycles 
corresponding to Figure \ref{figure-h}.

For $N$ cycles, 
the noise contribution $n(t)$ 
can be reduced effectively 
to $n_{eff}(t) \equiv 
\frac1N \sum_a n_a(t) 
\sim \bar{n}/\sqrt{N}$, 
when the noise obeys a Gaussian distribution, 
we denote $n_a(t) \equiv n(t + (a-1)T_P)$ 
and $N$ is large. 
Namely, $n_{eff}(t)$ gets smaller 
$\propto N^{-1/2}$, 
as $N$ increases. 
In Figure \ref{figure-h}, roughly estimating, 
the typical size of $n_{eff}(t)$ is  
$\sim n(t)/1100, n(t)/1600, n(t)/2000, n(t)/2300$, 
respectively, 
for 
 $N \sim 1.3 \times 10^6, 2.6 \times 10^6, 3.9 \times 10^6, 5.2 \times 10^6$.  
These estimated noise contributions are  
consistent with Figure \ref{figure-h}.

\begin{figure}
\includegraphics[width=8.5cm]{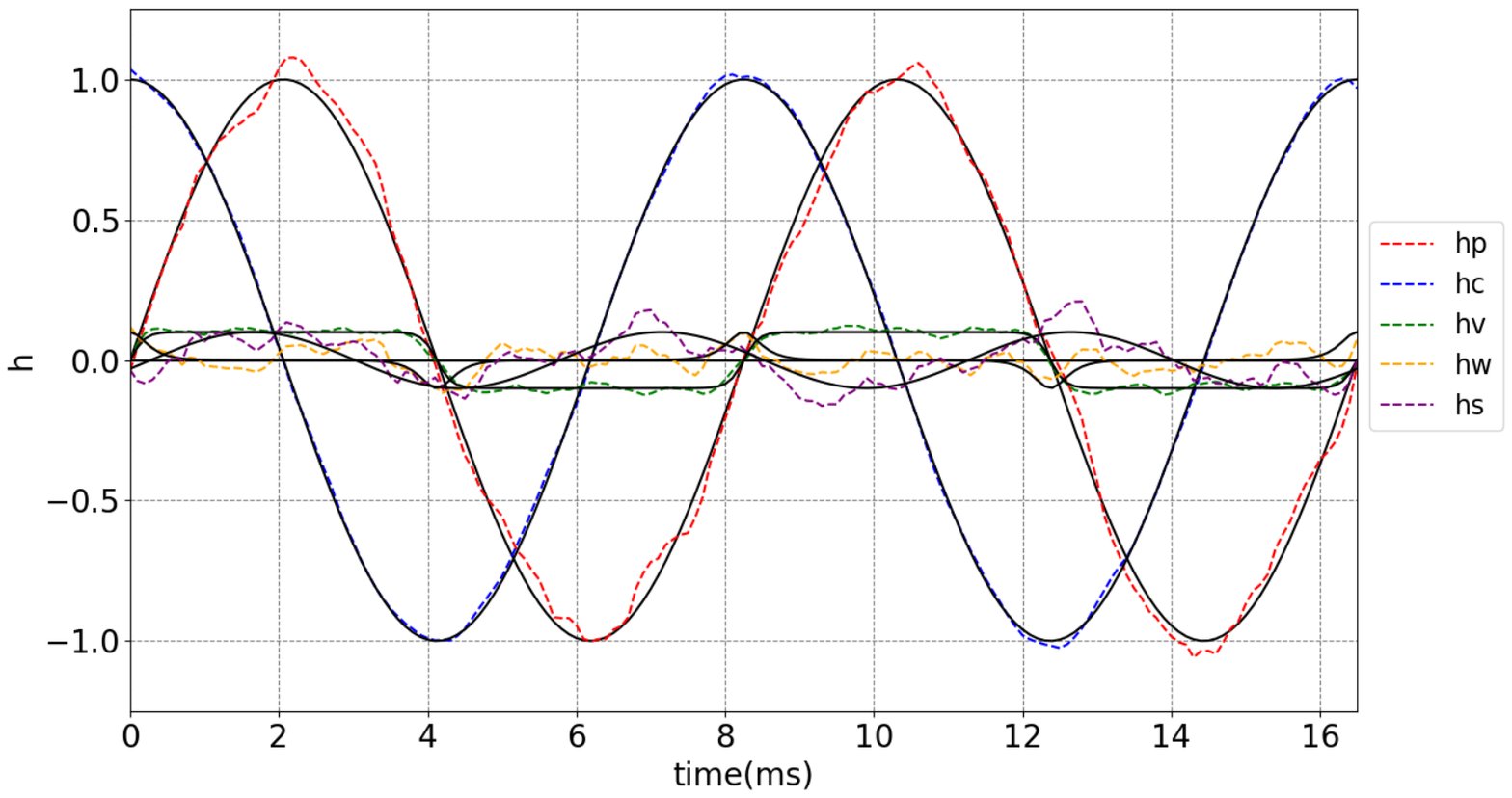}
\includegraphics[width=8.5cm]{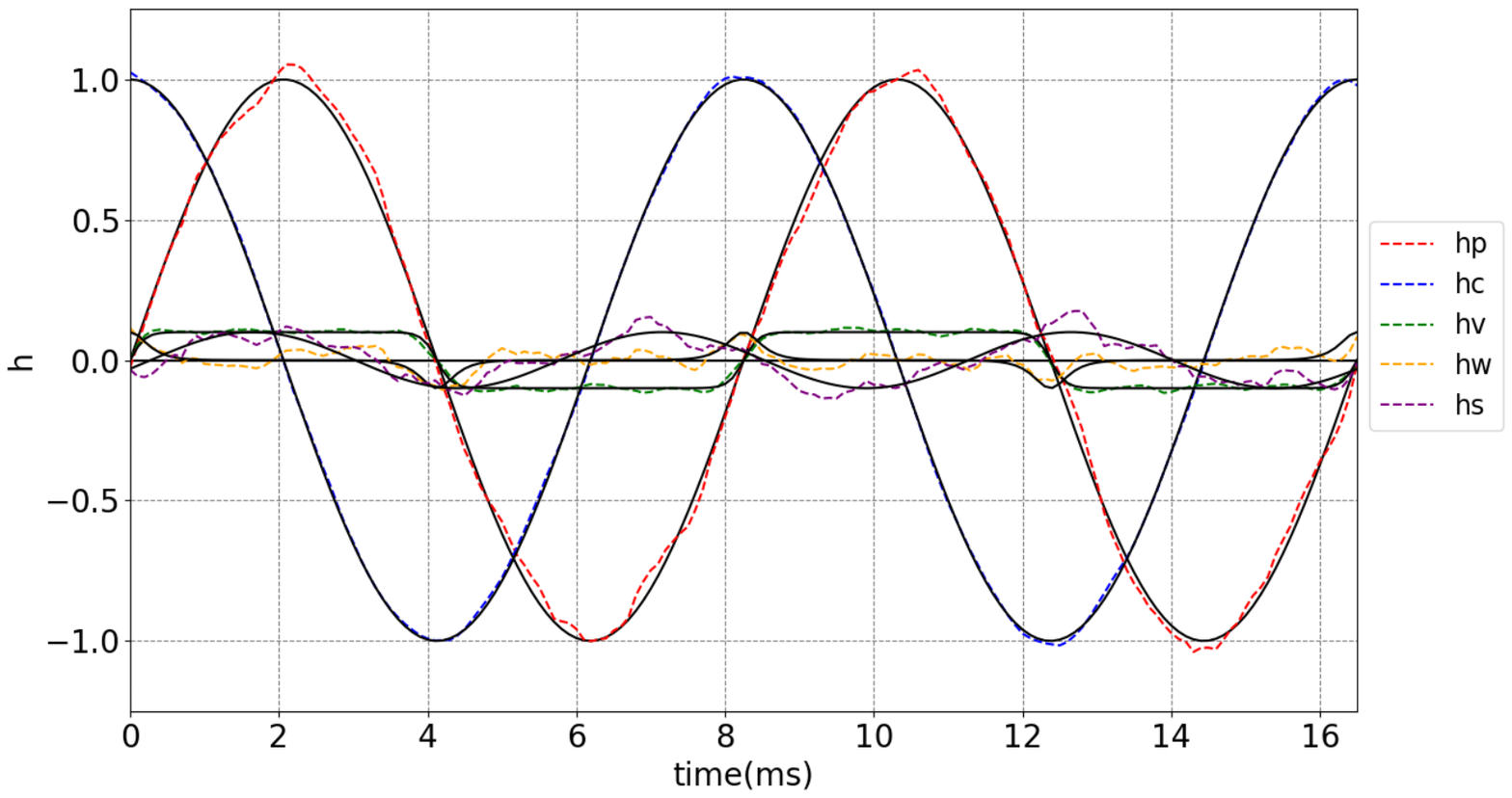}
\caption{
Time-domain reconstruction including non-sinusoidal waveforms: 
The detector location with its x-arm direction and the GW source position 
are the same as those in Figure \ref{figure-h}.  
$T_P$ is 16.5 msec.  
Top panel: 
$N$ is chosen as $5184000$, 
corresponding to nearly 24 hours. 
Bottom panel: 
$N = 10368000$ (48 hours). 
The unit of the vertical axis is arbitrary, 
normalized by the amplitude of the TT modes 
(blue or red plots in color). 
The amplitude of the S, V and W modes 
(purple, green or yellow plots in color) 
and 
the Gaussian noise 
are the same as those in Figure \ref{figure-h}. 
The delay of the S mode is 0.275 msec. 
The Jacobi elliptic functions $sn(t; k)$ and $cn(t; k)$ 
are assumed for $h_V(t)$ and $h_W(t)$, respectively, 
where the modulus $m = k^2$ is 0.9999999998. 
A sharp waveform in the Jacobi $cn$ function 
for W mode is reconstructed 
better for 48 hours than for 24 hours.}
\label{figure-Jacobi}
\end{figure}

Figure \ref{figure-Jacobi} shows another numerical reconstruction, 
where non-sinusoidal waveforms are mixed. 
The Jacobi elliptic $sn$ and $cn$ functions 
are assumed for $h_V(t)$ and $h_W(t)$, respectively. 
By using Eq. (\ref{sol}), 
each polarization state is reconstructed 
from strain outputs including not only sinusoidal but also 
non-sinusoidal small components. 
. 

\section{Future prospects and possible other effects}
In this section, we briefly discuss 
possible other effects on the current method and result.

\subsection{Stationary Gaussian noise}
First, 
the stationary Gaussian noise is assumed in Section III. 
The real noise is dependent on time. 
Regarding this issue, 
we can assume that the noise is still Gaussian but time-dependent 
where the standard deviation of the noise is denoted as $\sigma(t)$. 
For $N$ cycles, we denote 
$\sigma_a(t) \equiv \sigma(t + (a-1) T_P)$. 
For regressions in such a case, Eq. (\ref{A}) is modified as 
\begin{align}
  A(t) \equiv& 
  \sum_{a=1}^{N} 
  \frac{1}{[\sigma_a(t)]^2}
  \Bigl(S_{a}(t) - F^I_a(t) h_I(t)\Bigr)^{2} .
  \label{A2}
\end{align}
The expected waveform $\overrightarrow{H}_N(t)$ 
for the $N$ cycles is obtained in the same form as Eq. (\ref{sol}) 
but the replacement as $\sum_a \to \sum_a(1/[\sigma_a(t)]^2)$ 
must be done in the definitions of $L_N(t)$ and $M_N(t)$ 
by Eqs. (\ref{L}) and (\ref{M}), respectively.

\subsection{Noise reduction by increased cycles}
Secondly, 
the expected continuous GW signal is 
much smaller than 
a  current detector noise. 
Namely, $\bar{n} \gg h_{TT}$. 
A large $N$ is thus required. 
For three months ($\sim 100$ days) and twelve years for example, 
the effective $n_{eff}(t)$ becomes 
$n(t)/23000, n(t)/150000$, respectively,  
where $T_p$ is still assumed to be 16.5 msec. 
From a twelve-year observation, 
$n_{eff}(t)$ will get much smaller $\sim 10^{-5} \times\bar{n}$.

The third-generation detectors such as the CE and ET 
are aiming at the detector amplitude spectral density 
of $\sim 2-8 \times10^{-25} \mbox{Hz}^{-1/2}$ 
for a range of 10-500 Hz 
according to their white papers 
\cite{CE, ET, CE2}. 
For a twelve-year observation of 
a pulsar with $T_p \sim 10$ milliseconds 
(corresponding to $\sim 10^2$ Hz), 
$N$ is $\sim 4 \times 10^{10}$. 
CE and ET are thus expected to put stronger constraint 
on the extra polarization amplitudes than 
the current LIGO observations. 
However, the present paper is unable to estimate  
expected signal-to-noise ratios with the proposed formulation, 
because explicit waveform templates are not assumed 
in this paper.

\subsection{Earth rotation modulation}
A third comment is related with the second one. 
For a very long-time observation such as three months or twelve years, 
a simple periodic model is not sufficient 
\cite{Stairs}. 
In addition to the Earth rotation, 
we have to take account of the orbital motion of the Earth
as well as the geophysical disturbance. 
These effects do not affect $h_I(t)$ 
but modify a function of time for $F^I(t)$. 
Hence, the existence and uniqueness from Eq. (\ref{sol})
still hold, where $M(t)$ is calculated from 
the accordingly modified $F^I(t)$. 
The frequency modulation due to the Doppler effect 
by the Earth orbital motion is stronger 
by nearly two orders of magnitude 
than that of the Earth rotation. 

The Earth's rotational speed changes  
mainly owing to the tidal interaction 
\cite{Williams, Stephenson}. 
The length of day (LOD) is changing 
at the rate of $\dot{T}_E \sim 2$ milliseconds per century 
\cite{Williams, Stephenson}, 
whereas giant earthquakes make the LOD longer, 
e.g. by 6.8 microarcseconds 
owing to the 2004 Indian Ocean earthquake 
\cite{Nature}. 
The rate of change in the LOD, 
expressed 
as $\dot{T}_E/T_E$, 
is 
$\sim 2 \:\mbox{msec.}/\mbox{day}/\mbox{century}$. 

The changing LOC plays no role in $h_I(t)$, 
while it may affect calculations of $F_a^I(t)$. 
Therefore, we discuss how much the effect by the LOC modulation is.
 The change in the Earth's spin period makes an apparent shift 
of both the direction of the targeted pulsar 
 and the detector reference angle at each GW cycle. 

The angle around the Earth spin axis is denoted as 
$\Theta(t)$. 
The angular velocity of the Earth is written as 
\begin{align}
\frac{d\Theta(t)}{dt} 
= 
\frac{2\pi}{T_E(t)} . 
\label{angularvelocity}
\end{align}
By taking account of the change in the LOD, 
this is expanded around the initial time $t=0$ as 
\cite{Williams, Stephenson}
\begin{align}
\frac{d\Theta(t)}{dt} 
=& 
\left[\frac{2\pi}{T_E(t)}\right]_0 
- 2\pi 
\left[\frac{\dot{T}_E(t)}{(T_E(t))^2}\right]_0 t 
\notag\\
&+O\left(\left[\frac{(\dot{T}_E(t))^2}{(T_E(t))^3} \right]_0 t^2 \right),
\label{angularvelocity2}
\end{align}
where 
the dot denotes the time derivative and 
the subscript $0$ denotes the value at the initial time.

By using Eq. (\ref{angularvelocity2}), 
the total angle of the Earth rotation 
during the observation time $T_{obs}$ 
becomes 
 \begin{align}
\Theta(T_{obs}) 
&= 
\int_0^{T_{obs}}dt \frac{d\Theta(t)}{dt}
 \notag\\
 &= 
2\pi  \left[\frac{T_{obs}}{T_E(t)}\right]_0 
- \pi (T_{obs})^2 
\left[\frac{\dot{T}_E(t)}{(T_E(t))^2}\right]_0 
\notag\\
&+ O\left((T_{obs})^3\left[\frac{(\dot{T}_E(t))^2}{(T_E(t))^3}\right]_0 \right) . 
\end{align}
The first term in the right-hand side of the second line 
is the total rotation angle in the case of the constant rotation. 
The second term means the dominant correction 
due to the change in the LOD, 
denoted as $\Delta\Theta$. 
It is evaluated as
\begin{align}
|\Delta\Theta(T_{obs})|  
\sim\: 
&4 \times 10^{-5} 
\times 
\left(\frac{T_{obs}}{12 \mbox{year}} \right)^2
\notag\\
&\times
\left(\frac{\dot{T}_E/T_E^2}{2\: \mbox{msec.}/\mbox{day}^2/\mbox{century}}\right) , 
\label{DeltaTheta}
\end{align}
which is in the unit of radians. 
Hence, $|\Delta\Theta(T_{obs})|$ is $\sim$ 
a few arcseconds. 

Applying Eq. (\ref{DeltaTheta}) to the antenna pattern function, 
the corresponding correction to $F^I_a(t)$ 
is thus 
$|\Delta F_a^I(T_{obs})| 
\sim |\partial F_a^I/\partial \theta| 
\times 
|\Delta\theta(T_{obs})| 
\sim O(1) \times |\Delta\Theta(T_{obs})| 
\sim O(10^{-5})$, 
where we use 
$\Delta\theta(T_{obs}) \sim \Delta\phi(T_{obs}) 
\sim \Delta\psi(T_{obs}) \sim \Delta\Theta(T_{obs})$. 
Therefore, the effect due to the LOD modulation 
is smaller by three digits than that of the pulsar spin modulation 
that is estimated below as $O(10^{-2})$ for the Crab pulsar.

\subsection{Modulation of a pulsar spin period}
In order to modify Eq. (\ref{sol}), 
on the other hand, we may need to take account of 
the modulation in the pulsar spin period 
\cite{Jaranowski, Stairs}, 
which affects both the amplitude and period of the GWs 
\cite{Isi2015,Isi2017,Woan}.

There are several known pulsars 
for which the spin down rate is measured 
by radio observations. 
For the Crab pulsar for instance, 
its age is comparable to 
$\sim T_p/|\dot{T}_p|  \sim 10^3$ years. 
The change $\Delta T_p$ in the spin period 
for an observational duration $T_{obs}$ is 
$\Delta T_p \sim \dot{T}_p T_{obs}$, 
which means 
\begin{align}
\frac{|\Delta T_p|}{T_p} 
&\sim 
\left(\frac{|\dot{T}_p|}{T_p} \right) 
T_{obs} 
\notag\\
&\sim 
10^{-2} 
\left(
\frac{10^3 \mbox{year}}{T_p/|\dot{T}_p|}
\right)
\left(
\frac{T_{obs}}{12 \mbox{year}}
\right) ,
\label{DeltaT}
\end{align}
where we consider the Crab pulsar. 
The change in the pulsar spin period 
may not be negligible in a long-time observation. 
It is a few percents for 
twelve-year observations of the Crab pulsar.

For such a pulsar, 
we can estimate the spin period $T_p(t)$ 
from the form of 
$\dot{T}_p(t) = - \alpha [T_p(t)]^{-(n-2)}$ 
with a coefficient $\alpha$  and $n$ called the braking index, 
where $n = 1 \sim 3$ for most of observed isolated pulsars  
except for a newly born millisecond pulsar 
for which the GW radiation reaction term 
as $n = 5$ is thought to  
be dominant 
for the slow down of the pulsar spin. 
In reality, known pulsar's spin periods are available 
from radio measurements. 
As a practical procedure, thereby, 
Eq. (\ref{S}) may be modified as 
\begin{align}
S_1(t) &\equiv S(t), 
\notag\\
S_2(t) &\equiv S(t+T_2), 
\notag\\
&\cdots  
\notag\\
S_N(t) &\equiv S(t+T_N), 
\label{S2}
\end{align}
where we denote 
$T_1 \equiv T_p(0)$, $T_2 \equiv T_p(T_1)$, 
$T_3 \equiv T_p(T_1+T_2), \cdots,  
T_N \equiv T_p(T_1 + \cdots T_{N-1})$ 
to take account of the spin period evolution 
as $T_p(t)$.


Here, let us make an order-of-magnitude estimate 
of the amplitude evolution 
\cite{Woan}. 
We consider the GR polarization, 
because there are no established theories 
on the time evolution of non-GR polarization. 
In the quadrupole approximation in GR, 
$h_{TT} \propto T_p^{-2}$. 
Hence, we find 
$|\dot h_{TT}| \propto T_p^{-3} |\dot{T}_p|$. 
For the total observational time $T_{obs}$, 
the change in the amplitude of the GR mode is 
\begin{align}
\frac{|\Delta h_{TT}|}{h_{TT}}  
&\sim 
\frac{|\dot h_{TT}|}{h_{TT}} T_{obs} 
\notag\\
&\sim 
\frac{|\dot{T}_p|}{T_p} T_{obs} 
\notag\\
&\sim 
10^{-2} 
\left(
\frac{10^3 \mbox{year}}{T_p/|\dot{T}_p|}
\right)
\left(
\frac{T_{obs}}{12 \mbox{year}}
\right) ,
\label{Deltah}
\end{align}
where we use the age of the Crab pulsar 
$\sim T_p/|\dot{T}_p|  \sim 10^3$ years. 
If non-GR amplitudes also obey a power law 
$h_S \propto T_p^{-\beta}$, 
$h_V \propto T_p^{-\gamma}$, 
$h_W \propto T_p^{-\gamma}$ 
with positive indices $\beta$ and $\gamma$, 
the modulation of the non-GR amplitudes 
follows the order-of-magnitude estimate 
same as Eq. (\ref{Deltah}). 
Namely, 
the total change in the amplitude 
for twelve-year observations 
is a few percents, 
if the slow down rate is 
$\sim 10^{-3}/ \mbox{year}$. 
This suggests that 
the error in the reconstructed amplitude 
is a few percents 
e.g. for $\sim 12$ year observation of the Crab pulsar, 
when the amplitude modulation is ignored. 
In other words, 
the constant amplitude approximation 
in the reconstruction method 
is valid with $\sim$ a few percent accuracy. 

On the other hand, 
in order to incorporate the amplitude modulation 
in the waveform reconstruction, 
specific models of non-GR modes are needed, 
while the amplitude evolution of TT modes 
can be computed by using the quadrupole formula for instance.

About pulsar glitches, radio measurements are crucial. 
If a pulsar glitch does not change the spin period, 
the current  method can be applied as it is, 
while the data set only during the glitch may be removed 
in calculations for the waveform reconstruction. 
However, some giant pulsar glitches may make 
a significant change in the spin period. 
In this case, we have to spit the strain data stream into two sets; 
one data set before the glitch and the other set after the glitch. 
In the waveform reconstruction for the former data set, 
the pulsar spin period before the glitch should be used and 
it is given from radio measurements. 
For the latter data set, the spin period from radio measurement 
after the glitch should be used.

\subsection{Possible propagation speed test}
Next, 
we mention the speed of extra GW modes 
\cite{LIGO2019}. 
The possible arrival time difference between 
the TT and extra modes 
does not change the current discussion, 
because only the GW period matters  
but the time translation 
does not affect the $N$-cycle averaging. 

See Figure \ref{figure-h}, in which the arrival time of the S-mode 
is different from the other modes including the GR ones. 
The time-domain reconstruction method allows to 
constrain/measure the propagation speed of the extra polarizations, 
if they exist, as discussed below. 

Let $c_{K}$ and $c_{TT}$ denote the propagation speed of 
the extra polarization ($K = S, V, W$) 
and TT modes, respectively. 
We introduce a parameter $\delta_K$ to characterize 
the difference between $c_K$ and $c_{TT}$ by 
$c_K = c_{TT} (1 + \delta_K)$. 
For a pulsar at distance $d_P$, 
the arrival time difference $\Delta T_K$ becomes 
\begin{align}
\Delta T_K = \frac{d_P \delta_K}{c_{TT}} + O\left((\delta_K)^2\right) ,
\label{DeltaT}
\end{align}
where $c_{TT}$ can be measured from 
a comparison of the GW speed and the light velocity 
for merger events such as GW170817 
in multi-messenger astronomy.

By using the linearized version in $\delta_K$ of Eq. (\ref{DeltaT}), 
the upper bound on $\delta_K$ 
could be placed as 
\begin{align}
|\delta_K| 
=& 
\frac{c_{TT} | \Delta T_K|}{d_P} 
\notag\\
<& 1 \times 10^{-15} 
\left( \frac{|\Delta T_K|}{\delta t} \right)
\left( \frac{\delta t}{0.1 \mbox{msec.}} \right)
\left( \frac{1 \mbox{kpc}}{d_P} \right) 
\left( \frac{c_{TT}}{c} \right) , 
\label{delta}
\end{align}
if the arrival time difference is not detected. 
Here,   
$c_{TT}$ is almost equal to the speed of light $c$ 
\cite{GW170817},   
a pulsar is at $d_P \sim 1$ kpc,
the time resolution of the detector 
$\delta t$ 
limits the accuracy of measuring the arrival time difference 
($|\Delta T_K| < \delta t$ 
unless the arrival time difference is detected), 
and $\delta t$ is assumed 
$\sim 0.1$ msec. 
This time resolution is corresponding to sampling rates $\sim 10$ kHz, 
which is satisfied by the current LIGO sample rate as $16$ kHz.

\subsection{Computational procedure}
Before closing this section, we briefly address an issue on 
computational procedures and costs. 
We focus on the computations of Eq. (\ref{eq-vec}), where 
we assume that the pulsar spin modulation and 
the Earth rotation modulation are computed somewhere else 
and hence the computational costs of them are not discussed here. 
Before we solve Eq. (\ref{eq-vec}), 
we have to calculate $\vec{L}_N(t)$ and $M_N(t)$ 
that are defined by Eqs. (\ref{L}) and (\ref{M}), respectively. 
A point is that the calculations of them do not include 
any comparison between two different times, 
say 100 days or 10 years. 
Here, the total number of $S(t)$ and $F_a^I(t)$ equals to 
the total number of the data points 
$\sim 12\:\mbox{years} \times 16 \:\mbox{kHz}
\sim O(10^{12})$. 
for which a huge data storage is apparently required. 

The present computational method is as follows. 
Once the strain output for the $(n+1)$-th cycle is obtained, 
it is added to $\overrightarrow{L}_n(t)$ 
that is the summation 
from the $0$-th cycle to the $n$-th one. 
According to the ephemeris, 
the values of the antenna pattern $F_a^I(t)$ 
are calculated for the $(n+1)$-th cycle.

As mentioned above, the present method does not 
make a comparison of any quantities at two different times. 
Therefore, we do not need a huge storage 
for keeping the big data of $S(t)$. 
Once the GW observation is done for each cycle 
(e.g. the $(n+1)$-th cycle), 
the new output data $S_{n+1}(t)$ is added to $\overrightarrow{L}_n(t)$
by using Eq. (\ref{L}), 
such that we can obtain 
$\overrightarrow{L}_{n+1}(t)$. 
We continue this procedure until the $N$-th cycle, 
such that $\overrightarrow{L}_N(t)$ can be obtained. 

On the other hand, 
the strain data is not needed 
when computing $M_N(t)$ that is the sum from $a=1$ to $a=N$. 
This procedure is nothing but adding the $n$-th cycle 
to the earlier cycles. 
Once $F_a^I(t)$ is evaluated at $(n+1)$-th cycle, 
therefore, only the new $F_{n+1}^I(t)$ is 
used for computing $F_{n+1}^I(t) F_{n+1}^J(t)$ 
that is added to $M_n(t)$ until the $N$-th cycle. 
Then, we obtain $M_{N}(t)$. 

This additive operation is repeated until the $N$-th cycle, 
in which it is not necessary to keep the whole strain data 
and the antenna pattern values in the storage.

In this way, 
we obtain 
$\overrightarrow{L}_N(t)$ and $M_N(t)$ to arrive at Eq. (\ref{eq-vec}). 
Eq. (\ref{eq-vec}) is five linear equations for each time 
$t \in [0, T_p)$, 
where the time step is determined by the detector sampling rate. 
The number of the independent linear equations 
is thus estimated as 
$T_p$ multiplied by the sampling rate, 
e.g. 
$\sim 16.5\:\mbox{msec.} \times 16\:\mbox{kHz} 
\sim 3\times 10^2$ 
for the Crab pulsar and the LIGO sampling rate. 
Therefore, we solve 
$O(10^2)$ linear equations as Eq. (\ref{eq-vec}),  
for which computational costs are unlikely to be extremely high 
as follows.

\subsection{Computing costs}
Let us make a rough order-of-magnitude estimate
of computing costs. 
In order to numerically obtain Eq. (\ref{eq-vec}), 
we need compute numerically Eqs. (\ref{L}) and (\ref{M}) 
for $\overrightarrow{L}_N(t)$ 
and $M_N(t)$, respectively. 
The number of the GW strain data points $S_a(t)$ 
is $N$. 
It is $O(10^{12})$, where 
a data sampling rate is still assumed to be $\sim 10$ kHz 
for twelve-year observations. 
The number of computational steps in Eq. (\ref{L}) is 
$5\times N$. 
This is smaller than that in Eq. (\ref{M}), 
which defines a $5\times 5$ matrix,  
as $25 \times N$. 
Hence, Eq. (\ref{M}) is dominant in computational costs. 

The number of computational steps for $F_a^I(t)$ for 
each polarization at each data point is roughly $O(100)$, 
because $F_a^I(t)$ is a polynomial of sine (or cosine) functions 
and computational steps for a sine (or cosine) function 
are assumed to be $O(10)-O(50)$. 
Hence, the total steps for computing Eq. (\ref{M}) 
is calculated as 
$O(5\times 100 \times 25 \times N) = O(12500\times N)$, 
where five polarization states are still considered. 
This is $O(10^{16})$ for $N=O(10^{12})$ in twelve years. 

The performance of a current high-speed personal computer (PC) 
is roughly 100 GFLOPS (giga floating point operations per second). 
Therefore, the computational time for obtaining Eq. (\ref{M}) 
and hence Eq. (\ref{eq-vec}) is roughly equal to 
$O(10^{16})$ steps divided by 100 GFLOPS.  
This leads to $O(10^5)$ seconds, namely one core-day. 
If extra time in data transfer and so on is taken account of, 
the total computational time is roughly a few core-days.

\section{Conclusion} 
We considered a possible daily variation of antenna patterns 
for a ground-based GW detector due to Earth rotation. 
By defining the CAAM 
for continuous GWs from a known pulsar, 
we showed that distinct polarization states 
can be reconstructed in time domain 
from a given set of the strain outputs at a single detector. 
 
Constraining the propagation speed of 
extra polarization modes, if they 
coexist with the TT modes, 
was also discussed. 
We have to await significant progress in computational technology 
before the present method can be applied 
also for all-sky surveys of unknown pulsars, 
if the pulsar GW period and sky location are included as fitting parameters.

We discussed also possible effects 
due to the LOD modulation  
as well as a secular change in the pulsar period. 
Numerical simulations are needed, 
when we wish to take account of these effects accurately.  
It is left for future.

\begin{acknowledgments} 
We thank the anonymous referee for a lot of useful suggestions and comments 
on the earlier version of the manuscript. 
We are most grateful to Ken-ichi Oohara for his helpful comments on 
how to make a rough order-of-magnitude estimation of computing costs. 
We would like to thank Atsushi Nishizawa, 
Takahiro Tanaka, Kotaro Kyutoku and Hiroki Takeda
for fruitful conversations. 
We wish to thank Yousuke Itoh, Nobuyuki Kanda, Hideyuki Tagoshi 
and Seiji Kawamura for stimulating discussions. 
We thank Yuuiti Sendouda , Ryuichi Takahashi, 
Naoya Era, Yuki Hagihara, Daisuke Iikawa and Naohiro Takeda, 
Ryuya Kudo, Ryousuke Kubo, and Shou Yamahira 
for the useful conversations. 
This work was supported 
in part by Japan Society for the Promotion of Science (JSPS) 
Grant-in-Aid for Scientific Research, 
No. 20K03963 (H.A.),  
in part by Ministry of Education, Culture, Sports, Science, and Technology,  
No. 17H06359 (H.A.).  
\end{acknowledgments}


\begin{thebibliography}{99}
\bibitem{Einstein1916}
A. Einstein, 
Sitzungsber. Preuss. Akad. Wiss. Berlin (Math. Phys.) 
{\bf 1916}, 688 (1916). 
\bibitem{Einstein1918}
A. Einstein, 
Sitzungsber. Preuss. Akad. Wiss. Berlin (Math. Phys.) 
{\bf 1918}, 154 (1918). 
\bibitem{Will}
C. M. Will, Living Rev. Relativity, {\bf 17}, 4 (2014). 
\bibitem{LVK}
B. P. Abbott, 
Living. Rev. Relativ., {\bf 21}, 3 (2018).
\bibitem{KAGRA}
T. Akutsu, et al.,
Nature Astron., {\bf 3}, 35 (2019). 
\bibitem{Eardley}
D. M. Eardley, D. L. Lee, A. P. Lightman, R. V. Wagoner, and C. M. Will, 
Phys. Rev. Lett. {\bf 30}, 884 (1973).
\bibitem{Nishizawa2009}
A. Nishizawa, A. Taruya, K. Hayama, S. Kawamura, and M. A. Sakagami, 
Phys. Rev. D {\bf 79}, 082002 (2009). 
\bibitem{Hagihara2018}
Y. Hagihara, N. Era, D. Iikawa, and H. Asada, 
Phys. Rev. D {\bf 98}, 064035 (2018).  
\bibitem{Hagihara2019}
Y. Hagihara, N. Era, D. Iikawa, A. Nishizawa, and H. Asada, 
Phys. Rev. D {\bf 100}, 064010 (2019).  
\bibitem{Hagihara2020}
Y. Hagihara, N. Era, D. Iikawa, N. Takeda, and H. Asada, 
Phys. Rev. D {\bf 101}, 041501(R) (2020).  
\bibitem{PTEP}
M. Arimoto, et al.,
Prog. Theor. Exp. Phys. {\bf 2015}, 00000 (2021). 
\bibitem{KAGRA-2022}
R. Abbott, et al.,
arXiv:2203.01270.
\bibitem{LIGO2016}
B. P. Abbott et al. (Virgo and LIGO Scientific Collaborations), 
Phys. Rev. Lett. {\bf 116}, 221101 (2016). 
\bibitem{LIGO2017}
B. P. Abbott et al. (Virgo and LIGO Scientific Collaborations), 
Phys. Rev. Lett. {\bf 119}, 141101 (2017). 
\bibitem{GW170817}
B. P. Abbott, et al., 
Astrophys. J. Lett. {\bf 848}, L12 (2017);
B. P. Abbott, et al., 
Astrophys. J. Lett. 848, L13 (2017). 
\bibitem{LIGO2019}
B. P. Abbott et al. (Virgo and LIGO Scientific Collaborations), 
Phys. Rev. Lett. {\bf 123}, 011102 (2019).  
\bibitem{Hayama}
K. Hayama, and A. Nishizawa, Phys. Rev. D {\bf 87}, 062003 (2013).
\bibitem{Isi2015}
M. Isi, A. J. Weinstein, C. Mead, and M. Pitkin, 
Phys. Rev. D 91, 082002 (2015).
\bibitem{Isi2017}
M. Isi, M. Pitkin, and A. J. Weinstein, 
Phys. Rev. D {\bf 96}, 042001 (2017). 
\bibitem{Takeda}
H. Takeda, A. Nishizawa, Y. Michimura, K. Nagano, K. Komori, 
M. Ando, and K. Hayama, 
Phys. Rev. D {\bf 98}, 022008 (2018).
\bibitem{ST}
B. F. Schutz, and M. Tinto, 
Mon. Not. R. Astr. Soc. {\bf 224}, 131 (1987). 
\bibitem{GT}
Y. G\"ursel, and M. Tinto, 
Phys. Rev. D {\bf 40}, 3884 (1989). 
\bibitem{CYC}
K. Chatziioannou, N. Yunes, and N. Cornish, 
Phys. Rev. D {\bf 86}, 022004 (2012). 
\bibitem{PW}
E. Poisson, and C. M. Will, 
{\it Gravity}, (Cambridge Univ. Press, UK. 2014). 
\bibitem{Book-Creighton}
J. D. E. Creighton, and W. G. Anderson, 
{\it Gravitational-Wave Physics and Astronomy: 
An Introduction to Theory, Experiment and Data Analysis}, 
(Wiley, NY, 2011). 
\bibitem{Book-Maggiore}
M. Maggiore, 
{\it Gravitational Waves: Astrophysics and Cosmology}, 
(Oxford Univ. Press, UK, 2018).  
\bibitem{Jaranowski}
P. Jaranowski, A. Krolak, and B. F. Schutz, 
Phys.Rev.D {\bf 58}, 063001 (1998).
\bibitem{LIGO-pulsar-2017}
B. P. Abbott, et al., 
Phys. Rev. D {\bf 96}, 122006 (2017). 
\bibitem{LIGO-pulsar-2019}
B. P. Abbott, et al., 
Astrophys. J. {\bf 879}, 10, (2019).
\bibitem{LIGO-2111}
R. Abbott, et al., 
arXiv:2111.13106.
\bibitem{LIGO-2112}
R. Abbott, et al., 
arXiv:2112.10990.
\bibitem{LIGO-ApJ-2017}
B. P. Abbott, et al., 
Astrophys. J. {\bf 839}, 12 (2017).
\bibitem{LIGO-PRL-2018}
B. P. Abbott, et al., 
Phys. Rev. Lett. {\bf 120}, 031104 (2018).
\bibitem{LIGO-PRD-2019}
B. P. Abbott, et al., 
Phys. Rev. D {\bf 99}, 122002 (2019).
\bibitem{LIGO-PRD-2022}
R. Abbott, et al.,
Phys. Rev. D {\bf 105}, 022002, (2022). 
\bibitem{LIGO-ApJ-2021b}
R. Abbott, et al., 
Astrophys. J. {\bf 921}, 80 (2021). 
\bibitem{LIGO-PRD-2022b}
R. Abbott, et al., 
Phys. Rev. D {\bf 105}, 082005 (2022). 
\bibitem{LIGO-2201b}
R. Abbott, et al., 
arXiv:2201.10104.
\bibitem{LIGO-2204b}
R. Abbott, et al., 
arXiv:2204.04523.
\bibitem{Papa2019}
V. Dergachev, and M. A. Papa, 
Phys. Rev. Lett. 
{\bf 123}, 101101 (2019). 
\bibitem{Papa2021}
V. Dergachev, and M. A. Papa, 
Phys. Rev. D {\bf 104}, 043003 (2021). 
\bibitem{LIGO-PRD-2021}
R. Abbott, et al.,
Phys. Rev. D {\bf 104}, 082004 (2021).
\bibitem{LIGO-best-2022}
R. Abbott, et al.,
arXiv:2201.00697. 
\bibitem{Papa2202b}
V. Dergachev, and M. A. Papa, 
arXiv:2202.10598.
\bibitem{PRD-2021c}
R. Abbott, et al.,
Phys. Rev. D
{\bf 103}, 064017 (2021). 
\bibitem{ApJ-2022c}
R. Abbott, et al.,
Astrophys. J. {\bf 929}, L19 (2022).
\bibitem{Weisberg2016}
J. M. Weisberg, and Y. Huang, 
Astrophys. J. {\bf 829}, 55 (2016).
\bibitem{Kramer2021}
M. Kramer, et al.,
Phys. Rev. X {\bf 11}, 041050 (2021). 
\bibitem{ET}
M. Maggiore, et al.,
JCAP. {\bf 03}, 050 (2020). 
\bibitem{CE}
D. Reitze, et al., 
Bulletin of the AAS, {\bf 51}, 7 (2019). 
\bibitem{CE2}
M. Evans, et al., 
arXiv:2109.09882. 
\bibitem{Takeda2019}
H. Takeda, et al., 
Phys. Rev. D {\bf 100}, 042001 (2019). 
\bibitem{NumericalRecipe}
W. H. Press, S. A. Teukolsky, W. Vetterling, and B. P. Flannery, 
{\it Numerical Recipes, 3rd edition}, 
(Cambridge Univ. Press, UK, 2007).
\bibitem{Stairs}
I. H. Stairs, 
Living Rev. Relativity, {\bf 6}, 5 (2003).
\bibitem{Williams}
G. E. Williams,
Reviews of Geophysics, 
{\bf 38}, 37 (2000). 
\bibitem{Stephenson} 
F. R. Stephenson, L. V. Morrison, 
and C. Y. Hohenkerk, 
Proc. R. Soc. A. 
{\bf 472}, 20160404 (2016).
\bibitem{Nature}
M. Hopkin, 
https://doi.org/10.1038/news041229-6
\bibitem{Woan}
G. Woan, M. D. Pitkin, B. Haskell, D. I. Jones, 
and P. D. Lasky, 
Astrophys. J. Lett. {\bf 863}, L40, (2018).
\end{thebibliography}
\end{document}